# Weak Hadronic decays of Charmed Baryons and Current – Algebra Scheme

Neetika Sharma, PK Chatley
*Department of physics,*
*Dr. BR Ambedkar National Institute of Technology, Jalandhar*

Avinash C. Sharma*
*School of Basic & Applied Sciences,*
*GGS Indraprastha University, Delhi 110403, India*

The weak decays of charmed baryon multiplets into ground state baryons and mesons are investigated using current algebra technique. Some of the interesting results on the partial decays= rates and the asymmetry parameters for the $\Delta C=\Delta S=-1$ mode are calculated and consequences discussed.

In the present work, we have investigated the weak nonleptonic decays of B(3*) and B(3) charmed multiplets in the Cabbibo favoured mode using current algebra plus soft meson techniques, and assuming that the transitions are well approximated by 20"-dominance. We determine[1] the decay rates and the asymmetry parameters for the various decays and find that whereas the channel B(3) → B(6) + (P*) has vanishing asymmetries, the decays of particles $\Xi^{++}_2$ and $\Omega^{+}_2$ are completely forbidden in this channel.

We are aware that conventional GIM model ($H_w^{20"+84}$) has some unsatisfactory features in the charmed[2] as well as the uncharmed sectors[3]. In the uncharmed sector, a substantial. 15-admixture[4] can arise due to the incomplete cancellations

$$\left[\bar{d}\,\gamma_\mu\left(1+\gamma_\mu\right)\,u\,.\,\bar{u}\,\gamma_\mu\left(1+\gamma_\mu\right)\,s-\bar{d}\,\gamma_\mu\left(1+\gamma_\mu\right)\,c.\,\bar{c}\,\gamma_\mu\left(1+\gamma_\mu\right)s\right]$$

Owing to the large mass difference between u and c quarks, it has also been argued by Shiffman, *et. el*[5] that an additional 15-contribution can arise from soft gluon effects. In fact, the presence of 15 term in the weak Hamiltonian also helps in understanding the relation between pv and pc decays. Since 15 admixture does not contribute to Cabbibo enhanced decays, the pv decay amplitudes for the charmed baryons will be reduced effectively, because of small contribution from 20" part to $\Delta C = 0$, $\Delta S = 1$ decays. It would be eventually result in the revelation of the decay rates given in table 1. The numerical results are presented in the Tables below for the various modes.

We stress that at this stage it is difficult to make definite comments on the weak nonleptonic decays of charmed baryons and that, for the present, our results may only be of theoretical interest. Nevertheless, it will be useful to present our calculation in a definite numerical form of these may provide a clue for the analysis of charmed baryon decay data expected in future experiments.

## Table 1: Partial decay rates & asymmetry parameters

1(a) : B(3*) → B(8) + P(8)

| S.No. | Decay mode | Intermediate States | | Decay Amplitudes (in units of $10^{-6}$) | | Γ (in units of $10^{-10}$ MeV) | α |
|---|---|---|---|---|---|---|---|
| | | s-channel | u-channel | lAl | lBl | | |
| 1 | $\Lambda_1^{+'} \to \Lambda\pi^+$ | $\Sigma^+$ | $\Sigma_1^0$ | 0 | 0.799 | 0.063 | 0 |
| | $\to \Sigma^0\pi^+$ | $\Sigma^+$ | $\Sigma_1^0$ | 1.307 | 0.763 | 1.320 | -0.35 |
| | $\to \Sigma^+\pi^0$ | $\Sigma^+$ | $\Sigma_1^+$ | 1.307 | 0.763 | 1.320 | -0.35 |
| | $\to \Xi^0 K^+$ | $\Sigma^+$ | $\Xi_1^0, \Xi_1^{0'}$ | 0 | 3.852 | 0.456 | 0 |
| | $\to P\, K^0$/bar | $\Sigma^+$ | - | 0.9234 | 1.323 | 0.744 | -085 |
| 2 | $\Xi_1^{+'} \to \Xi^0\pi^+$ | - | $\Xi_1^0, \Xi_1^{0'}$ | 0.924 | 3.290 | 1.541 | -0.99 |
| | $\to \Sigma^+ K^0$/bar | - | $\Lambda_1^{+'}, \Sigma_1^+$ | 0.924 | 3.434 | 1.562 | -0.98 |
| 3 | $\Xi_1^{0'} \to \Xi^-\pi^+$ | $\Xi^0$ | - | 0.924 | 1.212 | 0.819 | 0.60 |
| | $\to \Xi^0\pi^0$ | $\Xi^0$ | $\Xi_1^0, \Xi_1^{0'}$ | 1.307 | 3.190 | 2.197 | -0.96 |
| | $\to \Sigma^+ K^-$ | $\Xi^0$ | $\Lambda_1^{+'}, \Sigma_1^+$ | 0 | 3.056 | 0.729 | 0 |
| | $\to \Lambda\, K^0$/bar | $\Xi^0$ | $\Sigma_1^0$ | 1.131 | 1.005 | 1.056 | -0.56 |
| | $\to \Sigma^0 K^0$/bar | $\Xi^0$ | $\Sigma_1^0$ | 0.653 | 0.266 | 0.326 | -0.26 |

1(b) : $B(3) \rightarrow B(6) + P(8)$

| S.No. | Decay mode | Intermediate States | | Decay Amplitudes (in units of $10^{-6}$) | | $\Gamma$ (in units of $10^{-10}$ MeV) | $\alpha$ |
|---|---|---|---|---|---|---|---|
| | | s-channel | u-channel | lAl | lBl | | |
| 1 | $\Xi_2^{+'} \rightarrow \Xi_1^0 \pi^+$ | $\Xi_1^+, \Xi_1^{+'}$ | - | 0 | 4.446 | 0.496 | 0 |
| | $\rightarrow \Xi_1^+ \pi^0$ | $\Xi_1^+, \Xi_1^{+'}$ | $\Xi_2^+$ | 0 | 3.142 | 0.248 | 0 |
| | $\rightarrow \Sigma_1^{++} K^-$ | $\Xi_1^+, \Xi_1^{+'}$ | - | 0 | 6.283 | 1.240 | 0 |
| | $\rightarrow \Sigma_1^+ K^0$/bar | $\Xi_1^+, \Xi_1^{+'}$ | - | 0 | 4.446 | 0.621 | 0 |
| | $\rightarrow P K^0$/bar | $\Xi_1^+, \Xi_1^{+'}$ | - | 0 | 6.2883 | 0.280 | 0 |
| 2 | $\Xi_2^{++} \rightarrow \Xi_1^+ \pi^+$ | - | $\Xi_2^+$ | 0 | 0 | 0 | 0 |
| | $\rightarrow \Sigma_1^{++} K^0$/bar | - | - | 0 | 0 | 0 | 0 |
| 3 | $\Omega_2^{+'} \rightarrow \Omega_1^0 \pi^+$ | $\Xi^0$ | - | 0 | 0 | 0 | 0 |
| | $\rightarrow \Xi_1^+ K^0$/bar | $\Xi^0$ | $\Xi_2^+$ | 0 | 0 | 0 | 0 |

1(c) : $B(3) \rightarrow B(3^*) + P(8)$

| S.No. | Decay mode | Intermediate States | | Decay Amplitudes (in units of $10^{-6}$) | | $\Gamma$ (in units of $10^{-10}$ MeV) | $\alpha$ |
|---|---|---|---|---|---|---|---|
| | | s-channel | u-channel | lAl | lBl | | |
| 1 | $\Xi_2^+ \rightarrow \Xi_1^{+'} \pi^0$ | $\Xi_1^+, \Xi_1^{+'}$ | $\Xi_2^+$ | 2.614 | 1.522 | 7.742 | -0.23 |
| | $\rightarrow \Xi_1^{0'} \pi^+$ | $\Xi_1^+, \Xi_1^{+'}$ | - | 1.848 | 0.218 | 3.824 | 0.05 |
| | $\rightarrow \Lambda_1^{+'} K^0$/bar | $\Xi_1^+, \Xi_1^{+'}$ | - | 1.848 | 0.218 | 3.802 | 0.05 |
| 2 | $\Xi_2^{++} \rightarrow \Xi_1^{+'} \pi^+$ | - | $\Xi_2^+$ | 1.848 | 2.345 | 4.053 | -0.48 |
| 3 | $\Omega_2^+ \rightarrow \Xi_1^{+'} K^0$/bar | - | $\Xi_2^+$ | 1.848 | 2.345 | 4.298 | -0.50 |

•

## Ackowledgments

The authors wish to thank Profs. MM Gupta and RC Verma for useful discussions.

## References

[1] The detailed formalism and the calculations are given in the Neetika Sharma, PK Chatley, A. Sharma, Preprint No. NIT-J/2008/Phys/03 (Submited to Phys.Rev. D-2008).

[2] G. Branco et al, Phys. Rev. D13, 680 (1976) ; V. Barger & S. Pakavasa, Phys. Rev, Lett. 43, 812 (1979).

[3] Y. Iwasaki, Phys. Rev. Lett. 34, 1407 (1975)

[4] Y. Igauashi & M. ShinMura, Nucl. Phys. B129 (1977); RC Verma, MP Khanna, Pramana, 99, 643 (1977).

[5] MA Shifman, AI Vainshteinm VI Zakharov, Nucl.Phys. B120, 16 (1977). ITEP Preprints ITEP-63 and 64 (1979).

* * * * * *